\documentclass[a4paper,11pt]{article}
\usepackage{pos}

\title{Composition Sensitivity for the Cosmic Ray Anisotropy with SWGO}
 \ShortTitle{Probing Cosmic Ray Anisotropy with SWGO}

\author*[a]{Andrew~M.~Taylor}
\author[b]{Gwenael Giacinti}
\author[c]{Paolo Desiati}
\author[c]{Juan Carlos Diaz Velez}
\author[d]{Andrea Chiavassa}
\author[e]{Guiseppe Di Sciascio}
\author[f]{Juan Carlo Arteaga Velazquez}
\author[b]{Samridha Kunwar}

\affiliation[a]{DESY, Platanenallee 6, 15738 Zeuthen, Germany}

\affiliation[b]{MPIK, Saupfercheckweg 1, 69117 Heidelberg, Germany}

\affiliation[c]{WIPAC, University of Wisconsin - Madison, 222 W. Washington Ave. Madison, WI 53703, U.S.A.}

\affiliation[d]{Dipartimento di Fisica, Universit\`a degli Studi di Torino, Torino, Italy}

\affiliation[e]{Istituto Nazionale di Fisica Nucleare, Sezione di Roma Tor Vergata, via della Ricerca Scientifica 1, 00133 Roma, Italy}

\affiliation[f]{Institute of Physics and Mathematics, Universidad Michoacana de San Nicol as de Hidalgo, Morelia, Mexico}

\forColl{SWGO}

\emailAdd{andrew.taylor@desy.de}
\emailAdd{Gwenael.Giacinti@mpi-hd.mpg.de}

\abstract{A number of cosmic-ray observatories have measured a change in both phase and amplitude of the dipole component in the distribution of cosmic-ray arrival directions above a primary energy of 100~TeV. We focus on probing the cosmic-ray dipole and multipole evolution in the energy region of mutli~TeV to beyond PeV with a future large-area gamma-ray observatory, such as the Southern Wide-field Gamma-ray Observatory (SWGO). The ability to discriminate between different mass groups is essential to understand the origin of this evolution. Through a consideration of the energy and mass resolution for cosmic-ray detection by such an observatory, we estimate its separation power for decomposing the full-particle anisotropy into mass groups. In particular, we explore the feasibility of probing the dipole evolution with rigidity with SWGO. In this way, we demonstrate the great potential that this instrument offers for providing a deeper understanding of the origin of the cosmic-ray anisotropy.}

\FullConference{37$^{\rm{th}}$ International Cosmic Ray Conference (ICRC 2021)\\
		July 12th -- 23rd, 2021\\
		Online -- Berlin, Germany}

\begin{document}
\maketitle

\section{Introduction}

One of the most direct probes into the origin of cosmic rays comes from studies of
their arriving anisotropy. Although these particles arrive largely isotropically,
the level of their underlying dipole and multipoles have in recent years started to
be probed \cite{Aartsen:2018ppz}. In particular, a large phase swing of the dipole
direction, and a corresponding dip in the overall dipole amplitude, has been observed 
in the preceding two energy decades below the knee. The origin of this feature, however, 
remains unclear.

\begin{figure}[h!]
\begin{center}
\includegraphics[angle=0,width=0.49\textwidth]{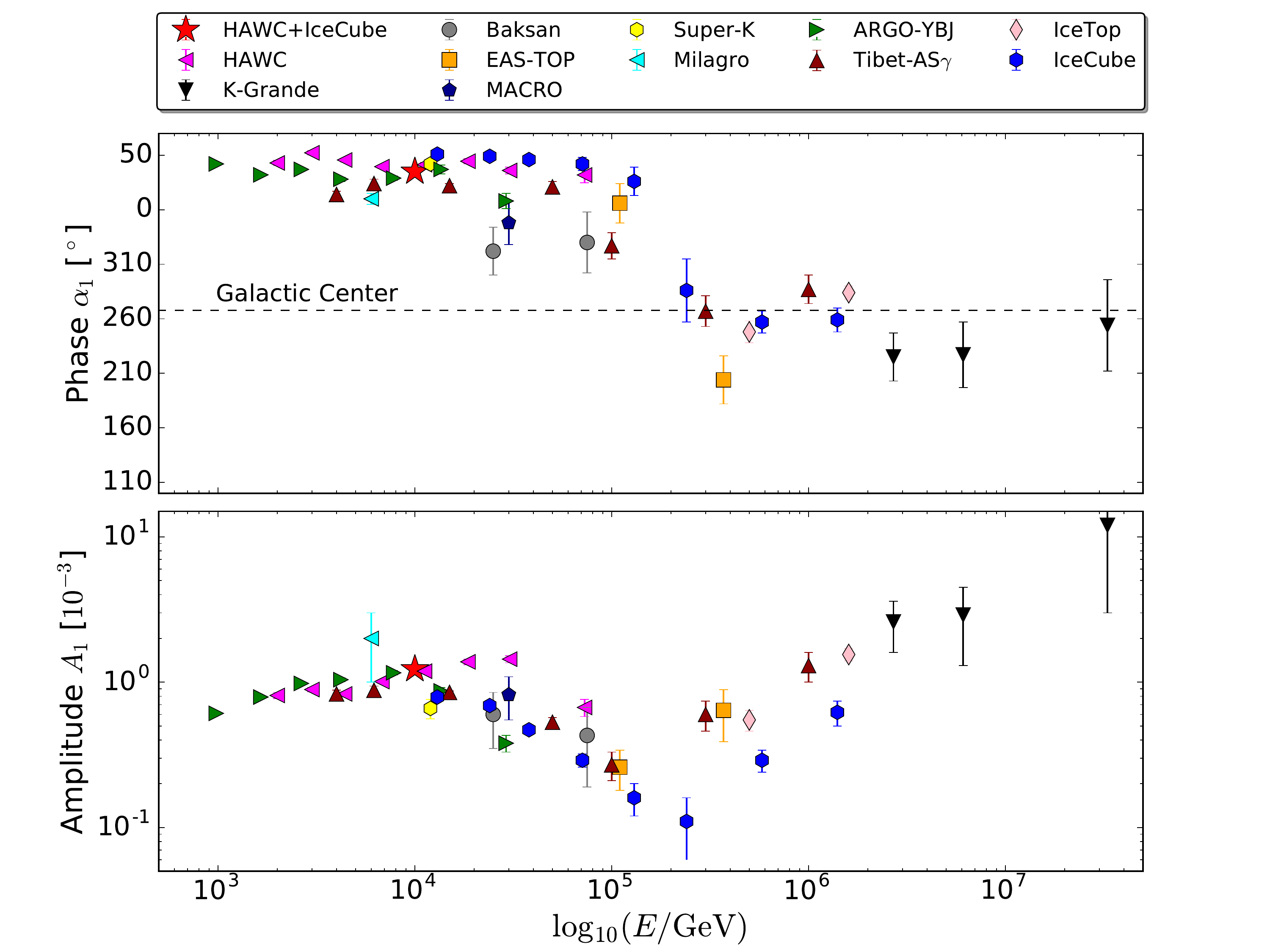}
\caption{The evolution in the phase (top-panel) and amplitude (bottom-panel) of the cosmic ray dipole, which exhibits a strong swing in the two energy decades below the knee energy scale. This plot has been taken from \cite{Aartsen:2018ppz}.}
\end{center}
\end{figure}

We here explore the role that the future detector SWGO, located in the southern
hemisphere on Earth, can play in providing a better understanding for the origin of 
this feature. Specifically, we focus on this instruments ability to separate out the
arriving cosmic rays into several major mass groups.

\section{Dipole Probe Energy Range}

\begin{figure}[h!]
\includegraphics[angle=0,width=0.49\textwidth]{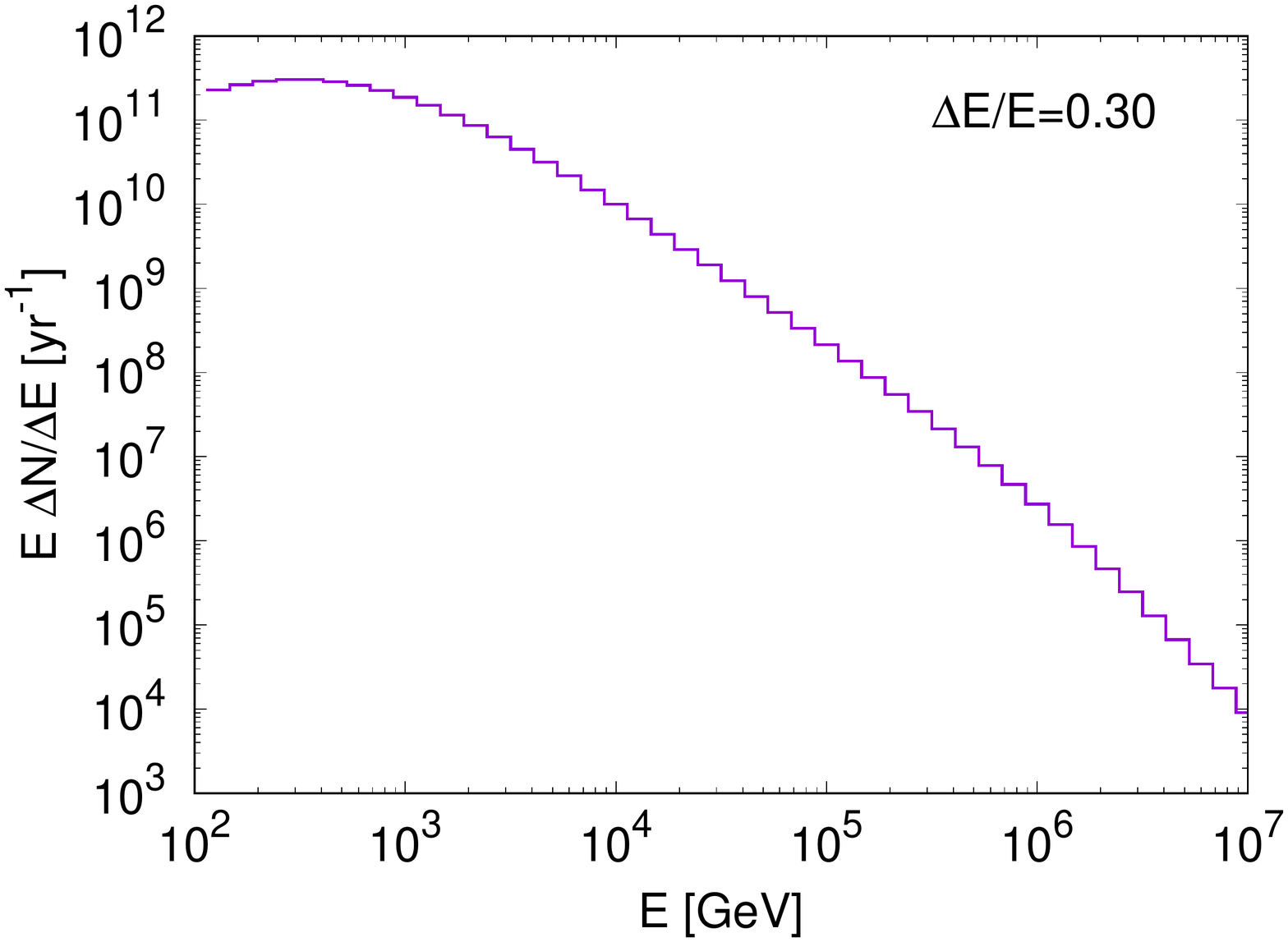}
\includegraphics[angle=0,width=0.49\textwidth]{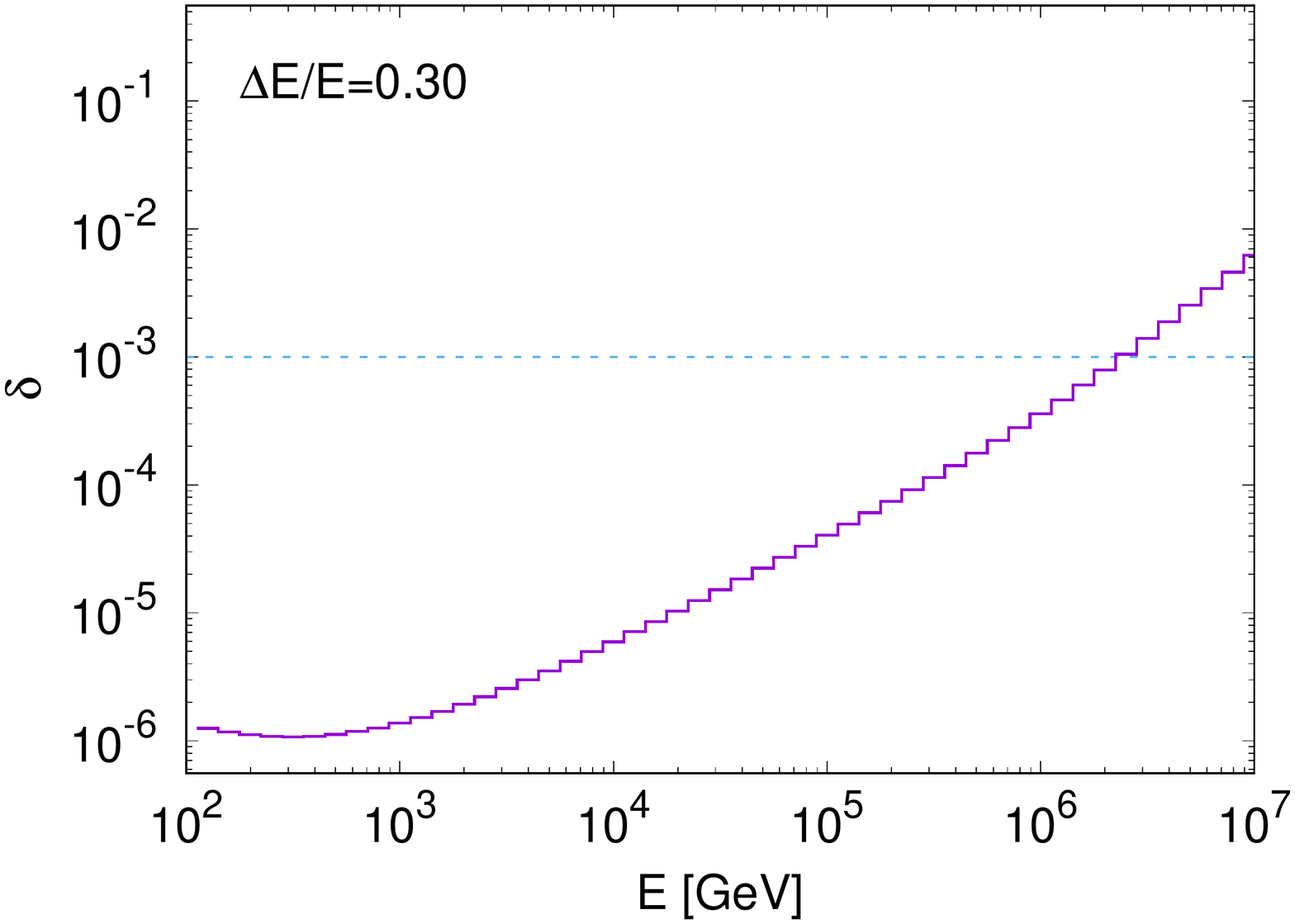}
\caption{Left-panel: the cosmic ray annual event rates as a function of energy determined for SWGO. Right-panel: the corresponding statistic limited level of anisotropy able to be probed as a function of energy (the energy at which the blue-dashed line and purple line intersect indicates the approximate energy limit that the dipole can be probed up to).}
\label{anisotropy_reach}
\end{figure}

A simple description for the effective area of SWGO, whose accuracy and validity
is sufficient over the energy range of consideration, can be found of the form \cite{Albert:2019afb}, can be approximately described by a smooth function of the form,
\begin{eqnarray}
A_{\rm Eff}(E)=A_{0}\left(\frac{E}{E_{0}}\right)^{4}\prod_{i=1,4}\left(\frac{1}{1+E/E_{i}}\right),
\end{eqnarray}
where $A_{0}=100~{\rm m}^{2}$, $E_{0}=10^{1.4}~{\rm GeV}$, $E_{1}=10^{1.7}~{\rm GeV}$, $E_{2}=10^{2.1}~{\rm GeV}$, $E_{3}=10^{2.5}~{\rm GeV}$, $E_{4}=10^{3}~{\rm GeV}$.
To obtain the rates of cosmic rays detectable by the instrument, this area is convolved 
with the cosmic ray spectrum below the knee, motivated by the spectral 
measurements in \cite{Apel:2013uni} of the form
\begin{eqnarray}
\frac{d\dot{N}}{dEdA} = 5 \times 10^{4} ~E_{\rm GeV}^{-2.7} ~e^{(-E_{\rm GeV} /10^{6} {\rm GeV})} ~{\rm m}^{2} ~{\rm s}^{-1} ~{\rm sr}^{-1} ~{\rm GeV}^{-1}
\end{eqnarray}
From this we obtain the expected annual rates of events expected in logarithmic energy bins of size $\Delta E/E$.
\begin{eqnarray}
E\frac{\Delta N}{\Delta E}=\Delta t_{\rm yr} \int_{E-\Delta E/2}^{E+\Delta E/2} \frac{d\dot{N}}{dEdA} A_{\rm eff}(E) dE
\end{eqnarray}
Using these rates the energy range that can probe a $10^{-3}$ level of anisotropy, assuming a full 
instrument year's worth of exposure, can then be determined. The left-hand panel of 
fig.~\ref{anisotropy_reach} shows the rates obtained, and the right-hand panel indicates 
the energy range of the anisotropy probe (note that the blue dashed line indicates the cosmic ray 
anisotropy level intended to be probed).

\section{Multipole Range to be Probed}
In order to estimate the noise power spectrum, we use the approximation from \cite{Ahlers:2016} that only depends on pixel-by-pixel Poisson noise and gives a flat spectrum $\mathcal{N}_\ell = \mathcal{N}$ given by
\begin{equation}\label{eq:noise}
\mathcal{N} \simeq \frac{1}{4\pi}\sum_i\frac{w_i^2\Delta\Omega^2}{\sum_\tau n_{\tau i}} \,,
\end{equation}
where $n_i$ is the number of events in pixel $i$ and $\Delta\Omega(\theta_{min},\theta_{max}) = 2\pi(\cos(\theta_{min})-\cos(\theta_{max}))$ is the solid angle of observed sky.

We can compare the expected angular power spectrum from the simple model in the form 
\begin{equation}\label{eq:ahlers}
C_\ell \propto 1/(2\ell +1)(\ell +2)(\ell+1)\,,
\end{equation}
introduced by \cite{Ahlers:2014}, scaled to the all-sky measurement in \cite{Aartsen:2018ppz} in order to estimate the highest observable multi-pole moment $\ell$. The corresponding relative uncertainty, assuming a Gaussian distribution of the individual $a_{\ell m}$’s, is estimated as 
\begin{equation}\label{eq:ahlerserr}
\langle \Delta C_\ell \rangle^2/C_\ell^2 \sim 2/(2\ell+1)/\textrm{fsky}\,,
\end{equation}
 where $\textrm{fsky}$ is the fraction of the sky observed. After one year, SWGO should be sensitive to structures larger than $\sim 35^\circ$ ($\ell = 5$) at $100~{\rm TeV}$ and larger than $\sim 10^\circ$ ($\ell = 20$) at $10~{\rm TeV}$.

\begin{figure}[h!]
\begin{center}
\includegraphics[angle=0,width=0.69\textwidth]{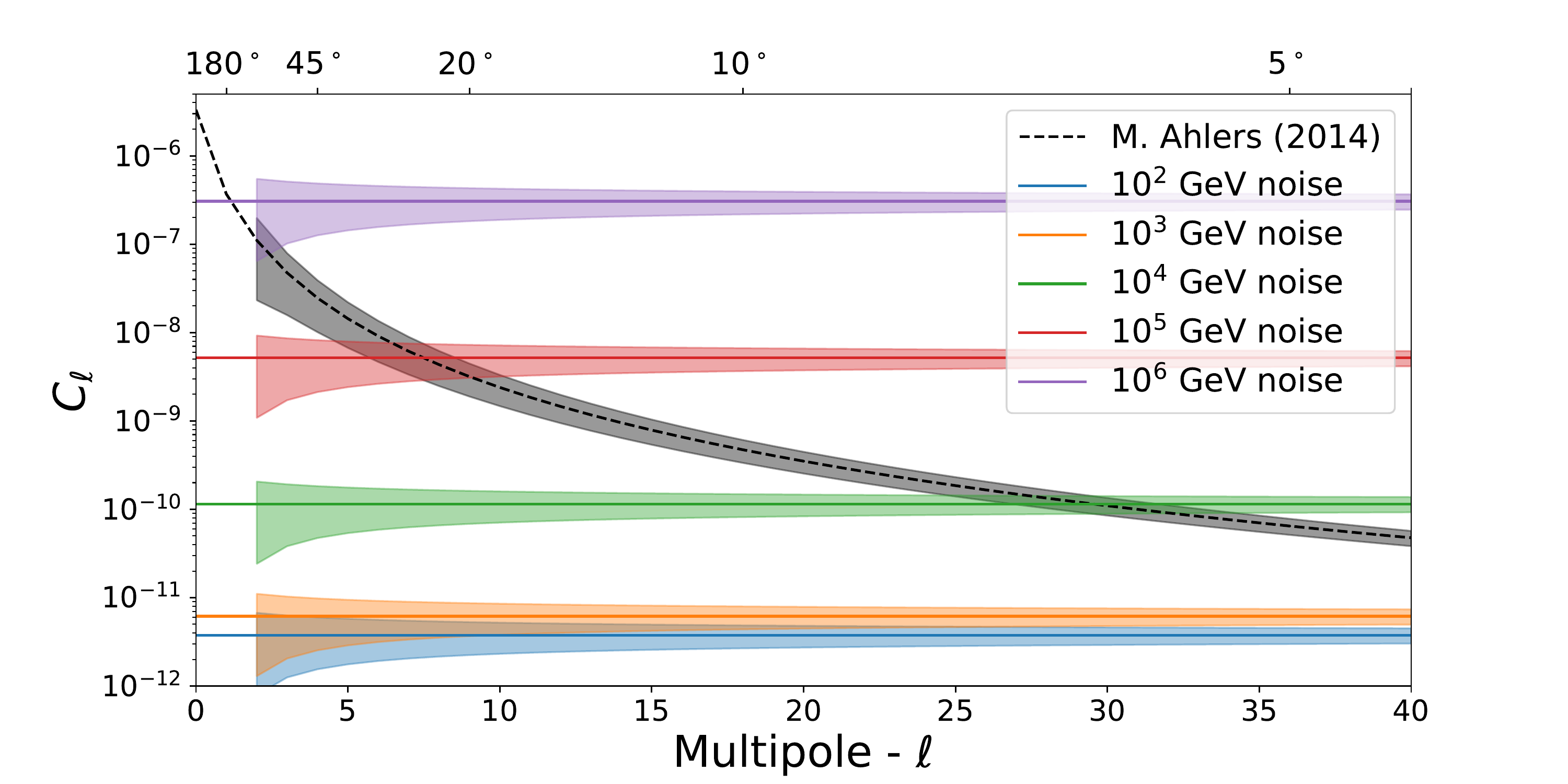}
\caption{Expected sensitivity to different angular scales for different energies after 1 year. Sensitivity is limited by statistics. The colored solid lines correspond to the angular power spectrum from Poisson noise calculated from Eq. \ref{eq:noise} at different energies, and dashed curve shows the model expectation for angular power spectrum from \cite{Ahlers:2014}.}
\end{center}
\end{figure}

\section{Separation Power into Mass Groups}

\begin{figure}[h!]
\begin{center}
\includegraphics[angle=0,width=0.49\textwidth]{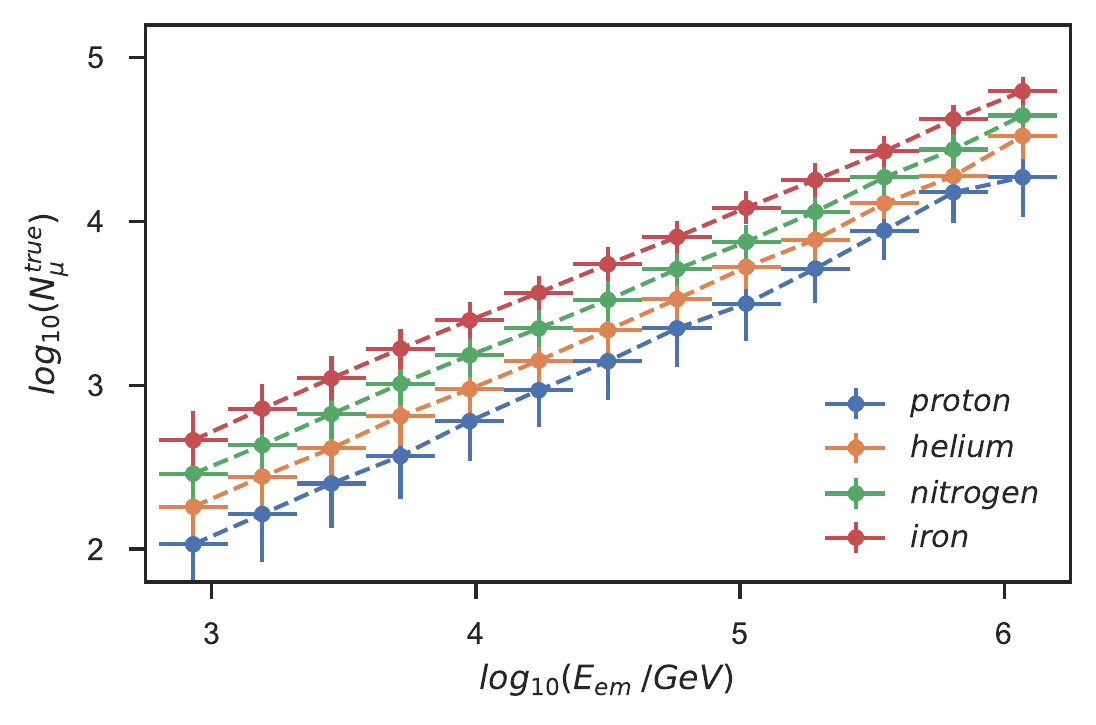}
\caption{Expected $\mu^\pm$ content of mass groups as a function of the electromagnetic energy deposited on the ground for particles with energy above 10 MeV simulated with -2 as slope of the primary energy for showers from $10^{3} \; {\rm to} \; 10^{6} \; {\rm GeV}$  and Zenith of from $0^\circ \; {\rm to} \; 65^\circ$ within 150 m from the shower core.}
\label{composition_separation}
\end{center}
\end{figure}

In order to simulate air showers, we use the CORSIKA 7.7400 simulation package~\cite{CORSIKA:1998aa}. For the standard simulated event set, we select the hadronic interaction model QGSJet-II.04~\cite{Optachenko:2011aa} for energies above 80~GeV. UrQMD 1.3.1~\cite{Bass:1998aa, Bleicher:1999aa} treats the low energy hadronic interactions. For electromagnetic processes, we use the EGS4 electromagnetic model~\cite{Nelson:1986aa}. Simulations were run for proton, helium, nitrogen and iron primary nuclei species, at an altitude of 4700~m with a spectral index of $-2$ for energy range from $10^{3} \; {\rm to} \; 10^{6} \; {\rm GeV}$ and Zenith angle from $0^\circ \; {\rm to} \; 65^\circ$. All ground level arriving muons within a radius of 150~m from the shower core were assumed to be counted by the detector. As indicated from fig.~\ref{composition_separation}, this level of muon counting as a function of the electromagnetic energy deposited on the ground for particles with energy above 10 MeV by the instrument would allow arriving cosmic ray showers to be separated into four distinct mass groups for primary cosmic ray energies above 1~TeV.

\section{Conclusion}

The proposed future instrument SWGO offers unique capabilities in terms of its anticipated large effective area, and good muon counting capabilities, and good energy resolution. Utilising these capabilities, our results 
indicate the great potential that this future instrument offers for cosmic ray anisotropy studies.
Specifically, our results indicate that this instrument is well suited for probing the origin
of the evolution of the dipole and multipoles of the arriving cosmic rays in the two energy decades
below the knee. Beyond sufficient cosmic ray statistics to probe this anisotropy, the separation power between several mass groups offers the potential to shed new light on the underlying physical origin of the anisotropy evolution with energy.

\end{document}